\documentclass[
letter
]{jpsj3}
\usepackage{txfonts}
\usepackage[dvipdfm]{color}

\newcommand{\TdegC}{\ensuremath{\mathrm{\char'27 \kern-.2em \hbox{C}}}}
\newcommand{\Tmum}{\ensuremath{ \mathrm{\mu m} }}
\newcommand{\Tsub}[2]{\ensuremath{#1_{\mathrm{#2}}}}
\newcommand{\TT}[1]{\Tsub{T}{#1}}
\newcommand{\Tkai}[1]{\Tsub{\chi}{#1}}

\newcommand{\TLMCO}[5]{La$_{#2}${#1}$_{#3}$Cu$_{#4}$O$_{#5}$}
\newcommand{\TMMCO}[6]{{#1}$_{#3}${#2}$_{#4}$Cu$_{#5}$O$_{#6}$}
\newcommand{\TMMCMO}[8]{{#1}$_{#4}${#2}$_{#5}$Cu$_{#6}${#3}$_{#7}$O$_{#8}$}
\newcommand{\TLECO}{\TLMCO{Eu}{1.8}{0.2}{}{4}}
\newcommand{\TLSCO}{\TLMCO{Sr}{2-x}{x}{}{4}}
\newcommand{\TNCO}{\TMMCO{Nd}{}{2}{}{}{4}}
\newcommand{\TNCCO}{\TMMCO{Nd}{Ce}{2-x}{x}{}{4}}
\newcommand{\TPCO}{\TMMCO{Pr}{}{2}{}{}{4}}
\newcommand{\TLECMO}{\TMMCMO{La}{Eu}{$M$}{1.8}{0.2}{1-y}{y}{4}}
\newcommand{\TLSCMO}{\TMMCMO{La}{Sr}{$M$}{2-x}{x}{1-y}{y}{4}}
\newcommand{\TNCCMO}{\TMMCMO{Nd}{Ce}{$M$}{2-x}{x}{1-y}{y}{4}}
\newcommand{\TLECNiO}{\TMMCMO{La}{Eu}{Ni}{1.8}{0.2}{1-y}{y}{4}}
\newcommand{\TNCCNiO}{\TMMCMO{Nd}{Ce}{Ni}{2-x}{x}{1-y}{y}{4}}

\newcommand{\TmSR}{\ensuremath{ \mu \mathrm{SR} }}

\newcommand{\TRevA}[1]{#1}
\newcommand{\TRevB}[1]{#1}
\begin{document}

\title{Pairing Symmetry Studied from Impurity Effects in the Undoped Superconductor T'-La$_{1.8}$Eu$_{0.2}$CuO$_4$}

\author{
Koki Ohashi$^1$, 
Takayuki Kawamata$^1$\thanks{E-mail: tkawamata@teion.apph.tohoku.ac.jp}, 
Tomohisa Takamatsu$^1$, 
Tadashi Adachi$^2$, \\
Masatsune Kato$^1$, 
and 
Yoji Koike$^1$
}

\inst{
$^1$Department of Applied Physics, Tohoku University, Sendai 980-8579, Japan, \\
$^2$Department of Engineering and Applied Sciences, Sophia University, Tokyo 102-8554, Japan
}

%%%%%%%%%%%%%%%%%%%%%%%%%%%%%%%%%%%%%%%%%%%%%%%%%%%%%%%%%%%%
\abst{
We have investigated the effects of magnetic Ni and nonmagnetic Zn impurities on the superconductivity in undoped T'-La$_{1.8}$Eu$_{0.2}$CuO$_4$ (T'-LECO) with the Nd$_2$CuO$_4$-type structure, using the polycrystalline bulk samples, to clarify the pairing symmetry. 
It has been found that both suppression rates of the superconducting transition temperature $T_\mathrm{c}$ by Ni and Zn impurities are \TRevA{nearly} the same and are very similar to those in the optimally doped and overdoped regimes of hole-doped T-La$_{2-x}$Sr$_{x}$CuO$_4$ with the K$_2$NiF$_4$-type structure. 
These results strongly suggest that the superconductivity in undoped T'-LECO is of the $d$-wave symmetry and is mediated by the spin fluctuation. 
}

\maketitle
\newpage
%%%%%%%%%%%%%%%%%%%%%%%%%%%%%%%%%%%%%%%%%%%%%%%%%%%%%%%%%%%%
%%%%%%%%%%%%%%%%%%%%%%%%%%%%%%%%%%%%%%%%%%%%%%%%%%%%%%%%%%%%
%\section{Introduction}
%%%%%%%%%%%%%%%%%%%%%%%%%%%%%%%%%%%%%%%%%%%%%%%%%%%%%%%%%%%%
It has long been believed that the high-\TT{c} superconductivity appears through the hole and electron doping into the antiferromagnetic mother compound \TMMCO{$Ln$}{}{2}{}{}{4} ($Ln$: lanthanide elements) with the K$_2$NiF$_4$-type (so-called T-type) and \TNCO-type (so-called T'-type) structure, respectively.\cite{ReviewSC} 
In electron-doped T'-\TMMCO{$Ln$}{Ce}{2-x}{x}{}{4}, it has also been known that the reduction annealing of the as-grown samples to remove excess oxygen occupying the apical site just above Cu in the CuO$_2$ plane is necessary for the appearance of superconductivity at $x \gtrsim 0.14$.\cite{Tokura:N337:1989:345,Radaelli:PRB49:1994:15322,Schultz:PRB53:1996:5157} 
Recently, in adequately reduced thin films of T'-{\TNCCO} (T'-NCCO), however, superconductivity has been observed in a wide range of $x$ \TRevB{and} even in the undoped mother compound of $x = 0$.\cite{Matsumoto:PC469:2009:924} 
Moreover, the superconductivity in undoped T'-\TMMCO{$Ln$}{}{2}{}{}{4} has been confirmed to appear in adequately reduced polycrystalline bulk samples of T'-\TMMCO{(La,Sm)}{}{2}{}{}{4} \cite{Asai:PC471:2011:682} and T'-{\TLECO} (T'-LECO) \cite{Takamatsu:APE5:2012:073101} obtained using low-temperature synthesis methods. 
In undoped T'-\TMMCO{$Ln$}{}{2}{}{}{4}, it has been suggested that carriers at the Fermi level are induced by the collapse of the charge-transfer gap between the upper Hubbard band of Cu3$d_{x^2 - y^2}$ and the O$2p$ band.\cite{Adachi:JPSJ82:2013:063713} 
Therefore, 
natures of the electronic state and superconductivity in undoped T'-\TMMCO{$Ln$}{}{2}{}{}{4} have attracted great interest. 
Although several studies of the transport, magnetic and optical properties have been carried out using thin films of superconducting (SC) undoped T'-\TMMCO{$Ln$}{}{2}{}{}{4},\cite{
Krockenberger:JJAP51:2011:010106,
Krockenberger:PRB85:2012:184502,
Krockenberger:SR3:2013:2235,
Kojima:PRB89:2014:180508(R),
Chanda:PRB90:2014:024503,
Schachinger:JPCM27:2015:045702,
ReviewNondope} 
the pairing symmetry of the superconductivity has not yet been clarified. 

The study of the effects of magnetic and nonmagnetic impurities on the superconductivity is useful to clarify the pairing symmetry of the superconductivity, 
because in a conventional $s$-wave superconductor, the SC transition temperature \TT{c} is suppressed by magnetic impurities more \TRevA{rapidly} than by nonmagnetic impurities. 
In an unconventional superconductor with a sign-changing SC gap such as a $d$-wave superconductor, 
on the other hand, 
\TT{c} is \TRevA{rapidly} suppressed by nonmagnetic impurities as well as by magnetic impurities.  
In fact, many studies of the impurity effect on \TT{c} have been carried out in high-\TT{c} cuprate superconductors.\cite{
Kang:PRB37:1988:5132,
Fujishita:SSC72:1989:529,
Tarascon:PRB42:1990:218,
Xiao:PRB42:1990:8752,
Felner:PC165:1990:247,
Yang:PRB42:1990:2231,
%Koike:SSC82:1992:889,
Harashina:PC212:1993:142,
Sreedhar:PCS227:1994:160,
Jayaram:PRB52:1995:3742,
Kluge:PRB52:1995:R727,
Brinkmann:PRB54:1996:6680,
Tallon:PRL79:1997:5294,
Nakano:PRB58:1998:5831,
Akoshima:PRB57:1998:7491,
%Kakinuma:PRB59:1999:1491,
%Kawamata:PRB62:2000:R11981,
Uchida:PC357:2001:25,
Jung:PRB65:2002:172501} 
In hole-doped T-{\TLSCO} (T-LSCO), 
%nonmagnetic impurities of Zn\TRevA{, if anything,} 
%suppress \TT{c} more than magnetic impurities of Ni.
\TRevB{the suppression of \TT{c} by nonmagnetic Zn is 
comparable or even slightly larger than the suppression by magentic Ni}.\cite{
%LSCO REF
Kang:PRB37:1988:5132,
Fujishita:SSC72:1989:529,
Tarascon:PRB42:1990:218,
Xiao:PRB42:1990:8752,
%Koike:SSC82:1992:889,
Harashina:PC212:1993:142,
Sreedhar:PCS227:1994:160,
Tallon:PRL79:1997:5294,
Nakano:PRB58:1998:5831}  
Such a tendency has also been observed in hole-doped 
Bi-2212 \cite{Kluge:PRB52:1995:R727,Akoshima:PRB57:1998:7491}
and 
YBa$_2$Cu$_3$O$_7$ \cite{Yang:PRB42:1990:2231,Harashina:PC212:1993:142,Tallon:PRL79:1997:5294}, 
leading to the conclusion that the pairing symmetry of hole-doped high-\TT{c} cuprate superconductors is unconventional $d$-wave \TRevB{mediated by} spin fluctuation.  
%However, the difference in the \TT{c} suppression between Ni and Zn impurities is not so large in T-LSCO. 
In electron-doped T'-NCCO, on the other hand, 
the \TT{c} suppression by Ni is much more \TRevA{rapid} than by Zn.\cite{%NCCO REF
Felner:PC165:1990:247,
Tarascon:PRB42:1990:218,
Sugiyama:PRB43:1991:10489,
Jayaram:PRB52:1995:3742} 
In the electron-doped so-called infinite-layer compound Sr$_{0.9}$La$_{0.1}$CuO$_2$, 
where the reduction annealing is necessary for the appearance of superconductivity as well as in T'-\TMMCO{$Ln$}{}{2}{}{}{4}, 
such a difference in the \TT{c} suppression between Ni and Zn impurities has also been observed.\cite{Jung:PRB65:2002:172501} 
Thus, hole-doped and electron-doped cuprate superconductors \TRevB{exhibited} a definite difference in the impurity effect on \TT{c}, suggesting 
%from the viewpoint of the impurity effect 
that the pairing symmetry of electron-doped high-\TT{c} cuprate superconductors 
\TRevB{might be} conventional $s$-wave. 
However, a number of experiments 
such as angle-resolved photoemission spectroscopy,\cite{Sato:S291:2001:1517,Matsui:PRL95:2005:017003} 
scanning superconducting quantum interference device (SQUID) \cite{Tsuei:PRL85:2000:182} 
and penetration depth measurements\cite{Kokales:PRL85:2000:3696} 
have revealed that electron-doped T'-\TMMCO{$Ln$}{Ce}{2-x}{x}{}{4} with $x = 0.15$ is regarded as a $d$-wave superconductor. 
Therefore, the impurity effect on \TT{c} especially 
in the electron-doped cuprate superconductors has not yet been understood. 

In this letter, to clarify the pairing symmetry of SC undoped T'-\TMMCO{$Ln$}{}{2}{}{}{4}, 
we have investigated the \TT{c} suppression by Ni and Zn impurities using polycrystalline bulk samples of T'-LECO. 
Surprisingly, it has been found that the obtained rates of the \TT{c} suppression by Ni and Zn impurities in T'-LECO are \TRevA{nearly} the same and are very similar to those in optimally doped and overdoped T-LSCO.\cite{%LSCO REF
Kang:PRB37:1988:5132,
Fujishita:SSC72:1989:529,
Tarascon:PRB42:1990:218,
Xiao:PRB42:1990:8752,
%Koike:SSC82:1992:889,
Harashina:PC212:1993:142,
Sreedhar:PCS227:1994:160,
Tallon:PRL79:1997:5294,
Nakano:PRB58:1998:5831}  
That is, the behavior is quite different from those of electron-doped T'-NCCO.\cite{
%NCCO REF
Felner:PC165:1990:247,
Tarascon:PRB42:1990:218,
Sugiyama:PRB43:1991:10489,
Jayaram:PRB52:1995:3742} 
These results suggest that 
the pairing symmetry of SC undoped T'-LECO is $d$-wave \TRevB{mediated by} spin fluctuation. 

%%%%%%%%%%%%%%%%%%%%%%%%%%%%%%%%%%%%%%%%%%%%%%%%%%%%%%%%%%%%
%\section{Experimental}
%%%%%%%%%%%%%%%%%%%%%%%%%%%%%%%%%%%%%%%%%%%%%%%%%%%%%%%%%%%%
Polycrystalline bulk samples of T'-{\TLECMO} ($M$ = Ni, Zn, and vacancy; $y = 0, 0.005, 0.01, 0.015$) were prepared by \TRevB{the low temperature synthesis method} 
described in detail in our former paper.\cite{Takamatsu:APE5:2012:073101} 
First, polycrystalline bulk samples of {\TLECMO} with the T-type structure were prepared by the conventional solid-state reaction method. 
Second, the solid-state reaction of T-{\TLECMO} and reductant CaH$_2$ was performed to obtain oxygen-reduced \TMMCMO{La}{Eu}{$M$}{1.8}{0.2}{1-y}{y}{3.5} 
with the \TMMCO{Nd}{}{4}{}{2}{7}-type structure.\cite{
%Chou:PRB42:1990:6172,
Pederzolli:JSSC136:1998:137,
Houchati:CM24:2012:3811} 
Third, T'-{\TLECMO} were obtained by heating \TMMCMO{La}{Eu}{$M$}{1.8}{0.2}{1-y}{y}{3.5} 
at 400 {\TdegC} for 12 h in a flowing O$_2$ gas. 
Finally, by the reduction annealing in vacuum at a pressure of $\sim 1 \times 10^{-4}$~Pa at $675\TdegC$ for 24~h to remove excess oxygen, 
SC polycrystalline bulk samples of T'-{\TLECMO} were obtained. 
The samples of $M$ = vacancy were prepared to \TRevA{examine} whether both Ni and Zn were 
well substituted for Cu in T'-LECO or not. 
In case neither Ni nor Zn were substituted for Cu in T'-LECO, 
the \TT{c} suppression by $M$ = Ni and Zn \TRevB{would be} \TRevA{nearly} the same as that by $M$ = vacancy. 
All the products were characterized by the powder x-ray diffraction using CuK$_\alpha$ radiation at room temperature. 
The analysis of the elemental distribution in the samples was performed 
by the scanning-electron-microscope-energy-dispersive-x-ray (SEM-EDX) spectroscopy. 
Magnetic susceptibility measurements were carried out 
%using a superconducting quantum interference device (SQUID) 
using a \TRevA{SQUID} 
magnetometer in a magnetic field of 10 Oe to estimate \TT{c}.

\begin{figure}
	\begin{center}
		\includegraphics[scale=0.9]{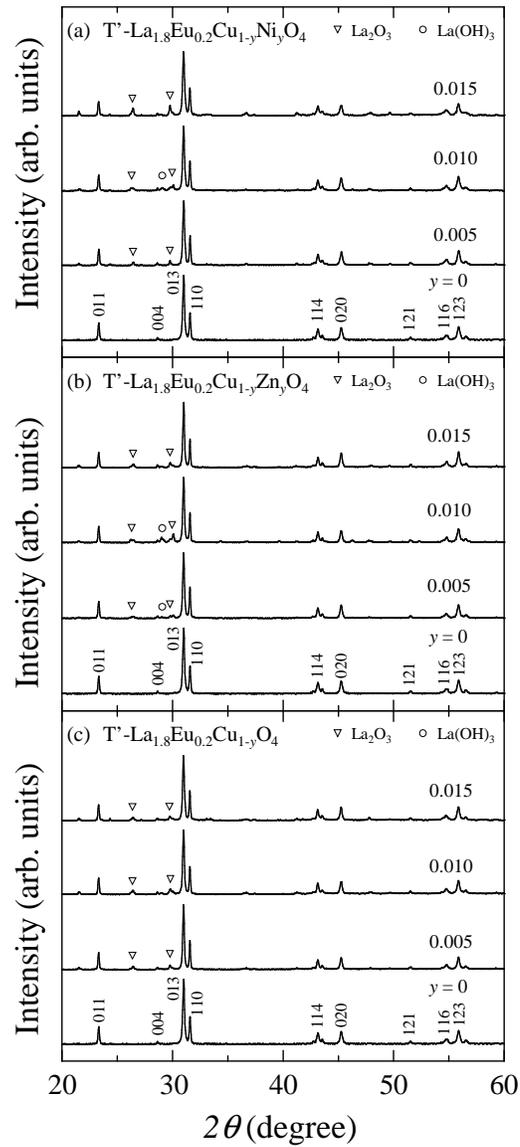}
		\caption{
Powder x-ray diffraction patterns using CuK$_\alpha$ radiation for T'-{\TLECMO} with (a) $M$ = Ni, (b) $M$ = Zn, and (c) $M$ = vacancy.
}
		\label{Fig:XRD}
	\end{center}
\end{figure}

\begin{figure}
	\begin{center}
		\includegraphics[scale=0.9]{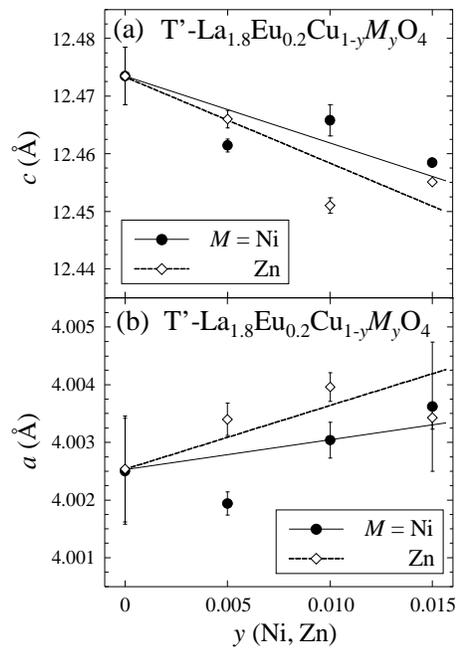}
		\caption{
Impurity-concentration $y$ dependence of the lattice constants (a) $a$ and (b) $c$ for T'-{\TLECMO} ($M$ = Ni, Zn). Solid and dashed lines are guides to the eyes. 
}
		\label{Fig:Latt}
	\end{center}
\end{figure}

%%%%%%%%%%%%%%%%%%%%%%%%%%%%%%%%%%%%%%%%%%%%%%%%%%%%%%%%%%%%
%\section{Results and Discussion}
%%%%%%%%%%%%%%%%%%%%%%%%%%%%%%%%%%%%%%%%%%%%%%%%%%%%%%%%%%%%
Figure \ref{Fig:XRD} shows powder x-ray diffraction patterns of the obtained samples of T'-{\TLECMO} ($M$ = Ni, Zn, and vacancy). 
It is found that all the samples are almost of the single phase, 
though there is a small amount of La$_2$O$_3$ and La(OH)$_3$ in the samples of $y \ge 0.005$. 
Figure \ref{Fig:Latt} shows the dependence on the Ni- and Zn-concentration $y$ of the lattice constants $a$ and $c$ determined by the powder x-ray diffraction. 
The obtained values of $a$ and $c$ of $y = 0$ are almost the same as those in our former paper.\cite{Takamatsu:APE5:2012:073101} 
Furthermore, it is found that the values of $a$ and $c$ monotonically increase and decrease with increasing $y$, respectively. 
These behaviors are similar to changes of $a$ and $c$ by the Ni and Zn substitution in T-LSCO,\cite{
Fujishita:SSC72:1989:529,
Xiao:PRB42:1990:8752} 
but are different from those in T'-NCCO with $x = 0$ \cite{Kurosaki:JAC350:2003:340} 
and $x = 0.15$.\cite{Tarascon:PRB42:1990:218} 
That is, the values of $a$ and $c$ in T'-{\TNCCMO} ($M$ = Ni, Zn) are independent of $y$.

The SEM-EDX spectroscopy of the obtained samples of T'-{\TLECMO} ($M$ = Ni, Zn) with $y$ = 0.01 has revealed that all the constituent elements are uniformly distributed 
without segregation of Ni and Zn within the spatial resolution of $1~\Tmum^2$. 

\begin{figure}
	\begin{center}
		\includegraphics[scale=0.8]{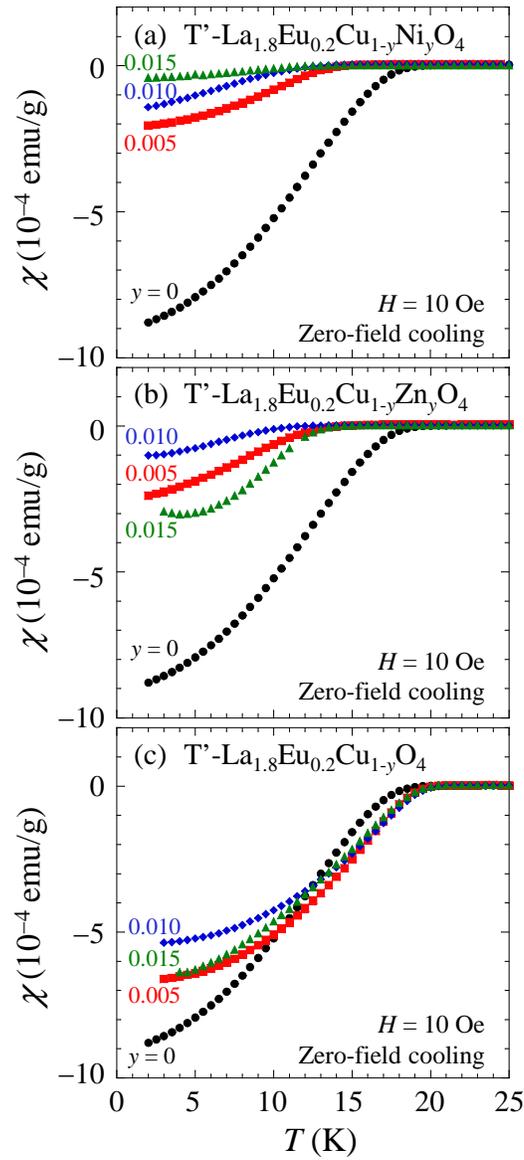}
		\caption{
(Color online) Temperature dependence of the magnetic susceptibility \Tkai{} 
of T'-{\TLECMO} with (a) $M$ = Ni, (b) $M$ = Zn, and (c) $M$ = vacancy measured 
in a magnetic field of 10 Oe on warming after zero-field cooling. 
}
		\label{Fig:Meiss}
	\end{center}
\end{figure}

\begin{figure}
	\begin{center}
		\includegraphics[scale=0.55]{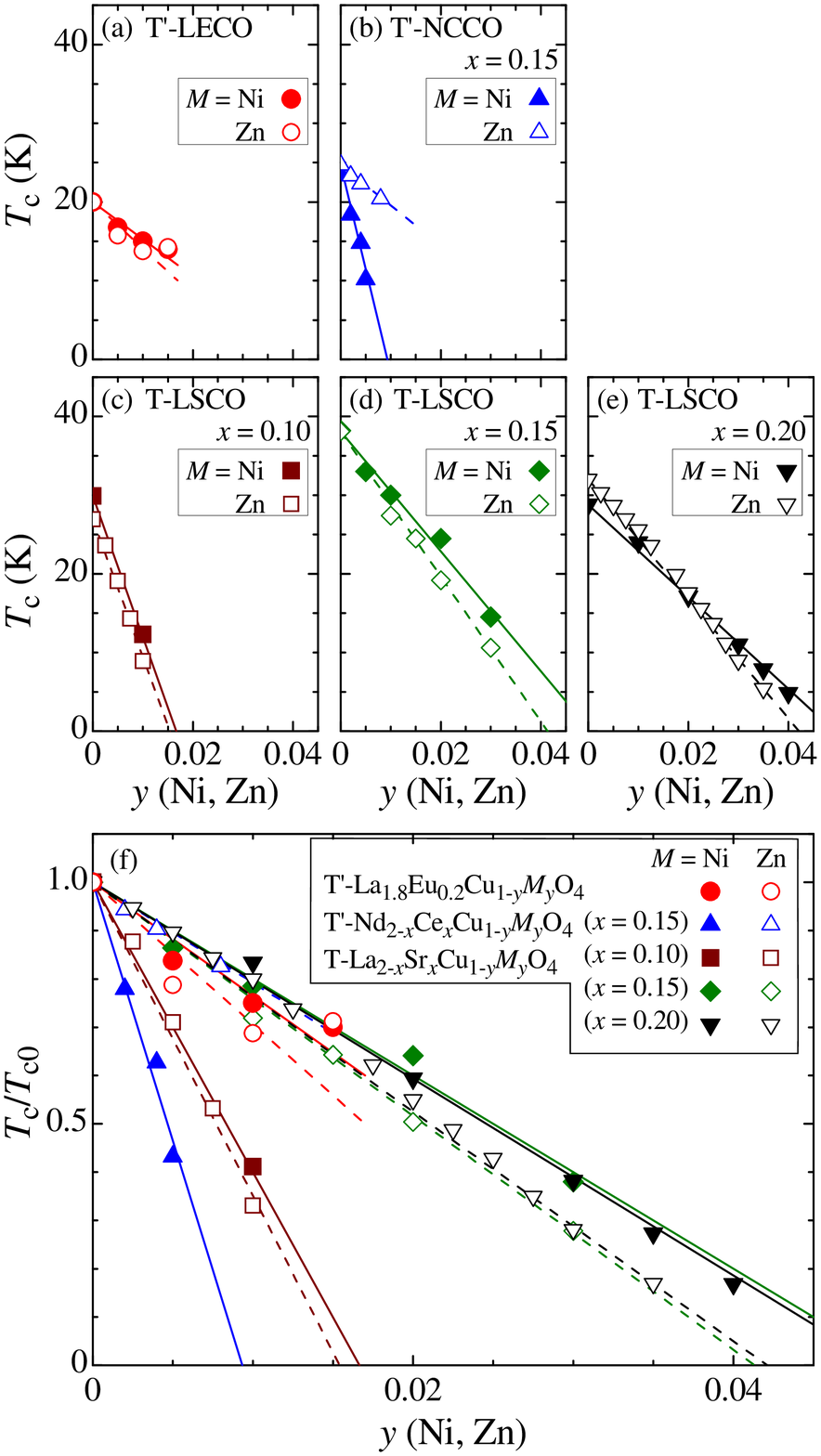}
		\caption{
(Color online) Impurity-concentration $y$ dependence of \TT{c} of 
(a) T'-{\TLECMO} ($M$ = Ni, Zn), 
(b) T'-{\TNCCMO} ($M$ = Ni, Zn),\cite{Tarascon:PRB42:1990:218} and 
T-{\TLSCMO} ($M$ = Ni, Zn) with 
(c) $x = 0.10$,\cite{Sreedhar:PCS227:1994:160,Uchida:PC357:2001:25} 
(d) $x = 0.15$,\cite{Tarascon:PRB42:1990:218} 
(e) $x = 0.20$.\cite{Sreedhar:PCS227:1994:160,Uchida:PC357:2001:25} 
(f) Impurity-concentration $y$ dependence of \TT{c}/\TT{c0} of these data, where \TT{c0} is \TT{c} at $y = 0$. 
Filled and open symbols indicate data of $M$ = Ni and Zn, respectively. 
Solid and dashed lines are guides to the eyes. 
}
		\label{Fig:Tc}
	\end{center}
\end{figure}

Figure \ref{Fig:Meiss} shows the temperature dependence of the magnetic susceptibility \Tkai{} measured 
%in a magnetic field of 10 Oe 
on warming after zero-field cooling for T'-{\TLECMO} ($M$ = Ni, Zn and vacancy). 
First, we discuss the behavior of $M$ = vacancy, to \TRevA{confirm that} both Ni and Zn are \TRevA{indeed} substituted for Cu in T'-LECO. 
It is found that both \TT{c} and the SC volume fraction 
are approximately independent of $y$ for $M$ = vacancy but clearly decrease 
with increasing $y$ for $M$ = Ni and Zn. 
This indicates that both Ni and Zn are successfully substituted for Cu in T'-LECO. 
Therefore, both decreases in \TT{c} and the SC volume fraction are due to the substitution of Ni and Zn for Cu. 
The decrease in the SC volume fraction by impurities has been observed in various hole-doped and electron-doped cuprate superconductors,\cite{
Kang:PRB37:1988:5132,
Tarascon:PRB42:1990:218,
%Koike:SSC82:1992:889,
Jung:PRB65:2002:172501}  
and explained in terms of the so-called Swiss cheese model 
due to the local destruction of superconductivity around impurities.\cite{Nachumi:PRL77:1996:5421} 
Figure \ref{Fig:Tc}(a) shows the $y$ dependence of \TT{c}, 
defined as the intersecting point between the extrapolated line of the steepest of the Meissner diamagnetism and zero susceptibility, 
for T'-{\TLECMO} ($M$ = Ni, Zn). 
Figures \ref{Fig:Tc}(b)--\ref{Fig:Tc}(e) show those of 
T'-{\TNCCMO} with $x = 0.15$,\cite{Tarascon:PRB42:1990:218} 
T-{\TLSCMO} with $x = 0.10$,\cite{Sreedhar:PCS227:1994:160,Uchida:PC357:2001:25} 
$0.15$,\cite{Tarascon:PRB42:1990:218} 
$0.20$,\cite{Sreedhar:PCS227:1994:160,Uchida:PC357:2001:25} respectively. 
%together with 
%those for T-{\TLSCMO} (
%$x = 0.10$ \cite{Sreedhar:PCS227:1994:160,Uchida:PC357:2001:25}, 
%$0.15$ \cite{Tarascon:PRB42:1990:218}, 
%$0.20$ \cite{Sreedhar:PCS227:1994:160,Uchida:PC357:2001:25}
%) 
%and T'-{\TNCCMO} ($x = 0.15$ \cite{Tarascon:PRB42:1990:218}). 
It is found that both suppression rates of \TT{c} in T'-LECO by the Ni and Zn substitution 
are \TRevA{nearly} the same. 
Surprisingly, this behavior is different from that of electron-doped T'-NCCO with $x = 0.15$ 
in spite of the same T'-type structure,\cite{%NCCO REF
Felner:PC165:1990:247,
Tarascon:PRB42:1990:218,
Sugiyama:PRB43:1991:10489,
Jayaram:PRB52:1995:3742} 
and is very similar to that of hole-doped T-LSCO 
with the different T-type structure.\cite{%LSCO REF
Kang:PRB37:1988:5132,
Fujishita:SSC72:1989:529,
Tarascon:PRB42:1990:218,
Xiao:PRB42:1990:8752,
%Koike:SSC82:1992:889,
Harashina:PC212:1993:142,
Sreedhar:PCS227:1994:160,
Tallon:PRL79:1997:5294,
Nakano:PRB58:1998:5831} 
Accordingly, this result suggests that the mechanism of the \TT{c} suppression by the Ni and Zn substitution in T'-LECO is similar to that in T-LSCO. 

In various high-\TT{c} cuprate superconductors, 
although the impurity effect on \TT{c} has been investigated,\cite{Uchida:PC357:2001:25} 
the mechanism of the \TT{c} suppression has not yet been understood \TRevB{systematically}. 
In this situation, 
\TRevA{it is a plausible understanding} 
%it has been reported 
that the \TT{c} suppression by impurities is due to the effect of carrier localization in the underdoped regime of hole-doped superconductors, 
while it is due to the pair-breaking effect based on the Abrikosov-Gorkov theory \cite{Abrikosov:SPJ12:1961:1243} 
in the overdoped regime.\cite{Kluge:PRB52:1995:R727} 
In hole-doped cuprate superconductors, in fact, 
the decrease in \TT{c}/\TT{c0}, where \TT{c0} is \TT{c} at $y = 0$, 
with increasing impurity-concentration in the optimally doped and overdoped regimes 
is smaller than that in the underdoped regime and is independent of the carrier concentration. 
To confirm this, the values of \TT{c}/\TT{c0} were plotted as shown in Fig. \ref{Fig:Tc}(f). 
It is found that the values of \TT{c}/\TT{c0} of T'-{\TLECMO} ($M$ = Ni, Zn) 
follow an universal line in the optimally doped and overdoped T-{\TLSCMO} ($M$ = Ni, Zn). 
In undoped T'-LECO, therefore, 
it follows that the \TT{c} suppression by the Ni and Zn substitution is due to the pair-breaking effect 
and that the SC pairing symmetry is $d$-wave \TRevB{mediated by} spin fluctuation 
\TRevA{as in the case of hole-doped T-LSCO}. 
In fact, a very recent {\TmSR} study has revealed that strong spin fluctuation coexists with 
the superconductivity in T'-LECO.\cite{Adachi:arXiv:1512.08095:2015} 
These results may imply that the \TRevA{electronic} state 
in undoped T'-LECO is very similar to that 
in the optimally doped and overdoped regimes of hole-doped T-LSCO, 
%, which is consistent with the decrease in \TT{c} by the hole doping 
%in T'-La$_{1.8-x}$Eu$_{0.2}A_x$CuO$_4$ ($A$ = Ca, Sr).
which \TRevB{may be} consistent with the result that \TT{c} decreases by the hole doping 
due to the Ca- or Sr-substitution for La 
in T'-LECO.\cite{Takamatsu:APE5:2012:073101,Takamatsu:PP58:2014:46} 

Considering Fig. \ref{Fig:Tc}(f), 
it is mysterious that only the \TT{c} suppression by the Ni substitution 
in electron-doped T'-NCCO with $x = 0.15$ is much more \TRevA{rapid} than that in undoped T'-LECO, 
because electron-doped T'-\TRevA{\TMMCO{$Ln$}{Ce}{2-x}{x}{}{4}} with $x = 0.15$ is regarded as a $d$-wave superconductor from a number of experiments 
\TRevA{as mentioned above}.\cite{Sato:S291:2001:1517,Matsui:PRL95:2005:017003,Tsuei:PRL85:2000:182,Kokales:PRL85:2000:3696} 
%such as angle-resolved photoemission spectroscopy,\cite{Sato:S291:2001:1517,Matsui:PRL95:2005:017003} 
%scanning SQUID \cite{Tsuei:PRL85:2000:182} 
%and penetration depth measurements.\cite{Kokales:PRL85:2000:3696} 
A very recent NMR study of SC T'-Pr$_{1.3-x}$La$_{0.7}$Ce$_x$CuO$_4$ with $x = 0.15$ 
has also revealed the existence of strong spin fluctuation, 
suggesting the $d$-wave paring.\cite{Yamamoto:JPSJ85:2016:024708} 
Furthermore, penetration depth measurements of the undoped T'-{\TPCO} thin film 
have also supported the $d$-wave paring.\cite{Chanda:PRB90:2014:024503} 
On the other hand, 
the strong \TT{c} suppression by magnetic impurities has also been observed 
in Fe- and Co-substituted T'-{\TNCCMO} ($M$ = Fe, Co) with $x = 0.15$ as well as in Ni-substituted T'-{\TNCCMO} with $x = 0.15$, 
\TRevB{which has been supposed to be an indication of} 
$s$-wave pairing.\cite{Tarascon:PRB42:1990:218,Sugiyama:PRB43:1991:10489,Jayaram:PRB52:1995:3742} 
A possible explanation for these contradictory results is that the strong \TT{c} suppression 
by the Fe, Co and Ni substitution is due to the insufficient removal of excess oxygen 
occupying the apical site just above Fe, Co and Ni substituted for Cu, 
because both Fe, Co and Ni tend to prefer the 6-fold octahedral 
coordination to the 4-fold square-planar one. 
Such a situation may take place also in Ni-substituted T'-{\TLECNiO}. 
\TRevA{However, T'-LECO has larger space around the apical site than T'-NCCO 
owing to larger La$^{3+}$ than Nd$^{3+}$ 
and moreover contains no Ce$^{4+}$ ions 
that more tightly binds excess oxygen than trivalent Ln$^{3+}$.} 
%Since the space around the apical site in T'-LECO is larger than in T'-NCCO 
%owing to the ionic radius of La$^{3+}$ larger than that of Nd$^{3+}$ and moreover Ce$^{4+}$ ions tend to attract excess oxygen more tightly than trivalent lanthanide ions, 
\TRevA{This may facilitate the removal of excess oxygen atoms in T'-{\TLECNiO}, 
which leads to a weaker \TT{c} suppression by the Ni substitution in T'-LECO 
than in T'-{\TNCCNiO} with $x = 0.15$.} 
This speculation may be supported by the above-mentioned result that the values of $a$ and $c$ change with $y$ in T'-{\TLECNiO} 
while they are independent of $y$ in T'-{\TNCCNiO} with $x = 0.15$.\cite{Tarascon:PRB42:1990:218} 
To be conclusive from the impurity effect on \TT{c}, further measurements using the same system of T'-\TMMCMO{$Ln$}{Ce}{$M$}{2-x}{x}{1-y}{y}{4} ($M$ = Ni, Zn) \TRevA{with various $Ln$ and} with various $x$ values from $x = 0$ to $0.15$ is necessary. 

%%%%%%%%%%%%%%%%%%%%%%%%%%%%%%%%%%%%%%%%%%%%%%%%%%%%%%%%%%%%
%\section{Summary}
%%%%%%%%%%%%%%%%%%%%%%%%%%%%%%%%%%%%%%%%%%%%%%%%%%%%%%%%%%%%
In \TRevB{summary}, 
we have investigated the \TT{c} suppression by the substitution of magnetic Ni and nonmagnetic Zn 
in SC undoped T'-LECO polycrystalline bulk samples 
to clarify the SC paring symmetry in undoped T'-LECO. 
From the magnetic susceptibility measurements, 
it has been found that both the Ni and Zn substitution suppress \TT{c} in the same degree. 
Surprisingly, this behavior is very similar to those in the optimally doped and overdoped regimes of hole-doped T-LSCO 
and is quite different from those in electron-doped T'-NCCO. 
The obtained $y$ dependences of \TT{c}/\TT{c0} in T'-{\TLECMO} ($M$ = Ni, Zn) are comparable to those in optimally doped and overdoped T-{\TLSCMO} ($M$ = Ni, Zn). 
These results strongly suggest that the SC pairing symmetry in undoped T'-LECO is $d$-wave, 
meaning that the pairing is medicated by the spin fluctuation. 
%This is the first study about the impurity effect on SC undoped T'-\TMMCO{$Ln$}{}{2}{}{}{4}. 

%%%%%%%%%%%%%%%%%%%%%%%%%%%%%%%%%%%%%%%%%%%%%%%%%%%%%%%%%%%%
%%%%%%%%%%%%%%%%%%%%%%%%%%%%%%%%%%%%%%%%%%%%%%%%%%%%%%%%%%%%

\begin{acknowledgment}
%\acknowledgment
This work was supported by JSPS KAKENHI Grant Number 23108004. 
\end{acknowledgment}

%%%%%%%%%%%%%%%%%%%%%%%%%%%%%%%%%%%%%%%%%%%%%%%%%%%%%%%%%%%%
%\newpage
%%%%%%%%%%%%%%%%%%%%%%%%%%%%%%%%%%%%%%%%%%%%%%%%%%%%%%%%%%%%

%%%%%%%%%%%%%%%%%%%%%%%%%%%%%%%%%%%%%%%%%%%%%%%%%%%%%%%%%%%%

%%%%%%%%%%%%%%%%%%%%%%%%%%%%%%%%%%%%%%%%%%%%%%%%%%%%%%%%%%%%
\bibliographystyle{jpsj}

\bibliography{taroRef}

\begin{thebibliography}{10}

\bibitem{ReviewSC}
For reviews, C. Chu, L. Deng, and B. Lv, Physica C \textbf{514}, 290 (2015); P.
  Fournier, Physica C \textbf{514}, 314 (2015).

\bibitem{Tokura:N337:1989:345}
Y.~Tokura, H.~Takagi, and S.~Uchida, Nature {\bfseries 337}, 345 (1989).

\bibitem{Radaelli:PRB49:1994:15322}
P.~G. Radaelli, J.~D. Jorgensen, A.~J. Schultz, J.~L. Peng, and R.~L. Greene,
  Phys. Rev. B {\bfseries 49}, 15322 (1994).

\bibitem{Schultz:PRB53:1996:5157}
A.~J. Schultz, J.~D. Jorgensen, J.~L. Peng, and R.~L. Greene, Phys. Rev. B
  {\bfseries 53}, 5157 (1996).

\bibitem{Matsumoto:PC469:2009:924}
O.~Matsumoto, A.~Utsuki, A.~Tsukada, H.~Yamamoto, T.~Manabe, and M.~Naito,
  Physica C {\bfseries 469}, 924 (2009).

\bibitem{Asai:PC471:2011:682}
S.~Asai, S.~Ueda, and M.~Naito, Physica C {\bfseries 471}, 682 (2011).

\bibitem{Takamatsu:APE5:2012:073101}
T.~Takamatsu, M.~Kato, T.~Noji, and Y.~Koike, Appl. Phys. Express {\bfseries
  5}, 073101 (2012).

\bibitem{Adachi:JPSJ82:2013:063713}
T.~Adachi, Y.~Mori, A.~Takahashi, M.~Kato, T.~Nishizaki, T.~Sasaki,
  N.~Kobayashi, and Y.~Koike, J. Phys. Soc. Jpn. {\bfseries 82}, 063713 (2013).

\bibitem{Krockenberger:JJAP51:2011:010106}
Y.~Krockenberger, H.~Yamamoto, M.~Mitsuhashi, and M.~Naito, Jpn. J. Appl. Phys.
  {\bfseries 51}, 010106 (2011).

\bibitem{Krockenberger:PRB85:2012:184502}
Y.~Krockenberger, H.~Yamamoto, A.~Tsukada, M.~Mitsuhashi, and M.~Naito, Phys.
  Rev. B {\bfseries 85}, 184502 (2012).

\bibitem{Krockenberger:SR3:2013:2235}
Y.~Krockenberger, H.~Irie, O.~Matsumoto, K.~Yamagami, M.~Mitsuhashi,
  A.~Tsukada, M.~Naito, and H.~Yamamoto, Sci. Rep. {\bfseries 3}, 2235 (2013).

\bibitem{Kojima:PRB89:2014:180508(R)}
K.~M. Kojima, Y.~Krockenberger, I.~Yamauchi, M.~Miyazaki, M.~Hiraishi, A.~Koda,
  R.~Kadono, R.~Kumai, H.~Yamamoto, A.~Ikeda, and M.~Naito, Phys. Rev. B
  {\bfseries 89}, 180508(R) (2014).

\bibitem{Chanda:PRB90:2014:024503}
G.~Chanda, R.~P. S.~M. Lobo, E.~Schachinger, J.~Wosnitza, M.~Naito, and A.~V.
  Pronin, Phys. Rev. B {\bfseries 90}, 024503 (2014).

\bibitem{Schachinger:JPCM27:2015:045702}
E.~Schachinger, G.~Chanda, R.~P. S.~M. Lobo, M.~Naito, and A.~V. Pronin, J.
  Phys.: Condens. Matter {\bfseries 27}, 045702 (2015).

\bibitem{ReviewNondope}
For a review, M. Naito, Y. Krockenberger, A. Ikeda, and H. Yamamoto, Physica C
  \textbf{523}, 28 (2016).

\bibitem{Kang:PRB37:1988:5132}
W.~Kang, H.~J. Schulz, D.~J{\'{e}}rome, S.~S.~P. Parkin, J.~M. Bassat, and
  P.~Odier, Phys. Rev. B {\bfseries 37}, 5132 (1988).

\bibitem{Fujishita:SSC72:1989:529}
H.~Fujishita and M.~Sato, Solid State Commun. {\bfseries 72}, 529 (1989).

\bibitem{Tarascon:PRB42:1990:218}
J.~M. Tarascon, E.~Wang, S.~Kivelson, B.~G. Bagley, G.~W. Hull, and R.~Ramesh,
  Phys. Rev. B {\bfseries 42}, 218 (1990).

\bibitem{Xiao:PRB42:1990:8752}
G.~Xiao, M.~Z. Cieplak, J.~Q. Xiao, and C.~L. Chien, Phys. Rev. B {\bfseries
  42}, 8752 (1990).

\bibitem{Felner:PC165:1990:247}
I.~Felner, D.~Hechel, and U.~Yaron, Physica C {\bfseries 165}, 247 (1990).

\bibitem{Yang:PRB42:1990:2231}
C.~Y. Yang, A.~R. Moodenbaugh, Y.~L. Wang, Y.~Xu, S.~M. Heald, D.~O. Welch,
  M.~Suenaga, D.~A. Fischer, and J.~E. Penner-Hahn, Phys. Rev. B {\bfseries
  42}, 2231 (1990).

\bibitem{Harashina:PC212:1993:142}
H.~Harashina, T.~Nishikawa, T.~Kiyokura, S.~Shamoto, M.~Sato, and K.~Kakurai,
  Physica C {\bfseries 212}, 142 (1993).

\bibitem{Sreedhar:PCS227:1994:160}
K.~Sreedhar, P.~Metcalf, and J.~Honig, Physica C: Superconductivity {\bfseries
  227}, 160 (1994).

\bibitem{Jayaram:PRB52:1995:3742}
B.~Jayaram, H.~Chen, and J.~Callaway, Phys. Rev. B {\bfseries 52}, 3742 (1995).

\bibitem{Kluge:PRB52:1995:R727}
T.~Kluge, Y.~Koike, A.~Fujiwara, M.~Kato, T.~Noji, and Y.~Saito, Phys. Rev. B
  {\bfseries 52}, R727 (1995).

\bibitem{Brinkmann:PRB54:1996:6680}
M.~Brinkmann, H.~Bach, and K.~Westerholt, Phys. Rev. B {\bfseries 54}, 6680
  (1996).

\bibitem{Tallon:PRL79:1997:5294}
J.~L. Tallon, C.~Bernhard, G.~V.~M. Williams, and J.~W. Loram, Phys. Rev. Lett.
  {\bfseries 79}, 5294 (1997).

\bibitem{Nakano:PRB58:1998:5831}
T.~Nakano, N.~Momono, T.~Nagata, M.~Oda, and M.~Ido, Phys. Rev. B {\bfseries
  58}, 5831 (1998).

\bibitem{Akoshima:PRB57:1998:7491}
M.~Akoshima, T.~Noji, Y.~Ono, and Y.~Koike, Phys. Rev. B {\bfseries 57}, 7491
  (1998).

\bibitem{Uchida:PC357:2001:25}
S.~Uchida, Physica C {\bfseries 357-360}, 25 (2001).

\bibitem{Jung:PRB65:2002:172501}
C.~U. Jung, J.~Y. Kim, M.-S. Park, M.-S. Kim, H.-J. Kim, S.~Y. Lee, and S.-I.
  Lee, Phys. Rev. B {\bfseries 65}, 172501 (2002).

\bibitem{Sugiyama:PRB43:1991:10489}
J.~Sugiyama, S.~Tokuono, S.~Koriyama, H.~Yamauchi, and S.~Tanaka, Phys. Rev. B
  {\bfseries 43}, 10489 (1991).

\bibitem{Sato:S291:2001:1517}
T.~Sato, T.~Kamiyama, T.~Takahashi, K.~Kurahashi, and K.~Yamada, Science
  {\bfseries 291}, 1517 (2001).

\bibitem{Matsui:PRL95:2005:017003}
H.~Matsui, K.~Terashima, T.~Sato, T.~Takahashi, M.~Fujita, and K.~Yamada, Phys.
  Rev. Lett. {\bfseries 95}, 017003 (2005).

\bibitem{Tsuei:PRL85:2000:182}
C.~C. Tsuei and J.~R. Kirtley, Phys. Rev. Lett. {\bfseries 85}, 182 (2000).

\bibitem{Kokales:PRL85:2000:3696}
J.~D. Kokales, P.~Fournier, L.~V. Mercaldo, V.~V. Talanov, R.~L. Greene, and
  S.~M. Anlage, Phys. Rev. Lett. {\bfseries 85}, 3696 (2000).

\bibitem{Pederzolli:JSSC136:1998:137}
D.~Pederzolli and J.~Attfield, J. Solid State Chem. {\bfseries 136}, 137
  (1998).

\bibitem{Houchati:CM24:2012:3811}
M.~I. Houchati, M.~Ceretti, C.~Ritter, and W.~Paulus, Chem. Mater. {\bfseries
  24}, 3811 (2012).

\bibitem{Kurosaki:JAC350:2003:340}
K.~Kurosaki, H.~Kobayashi, and S.~Yamanaka, J. Alloys Compd. {\bfseries 350},
  340 (2003).

\bibitem{Nachumi:PRL77:1996:5421}
B.~Nachumi, A.~Keren, K.~Kojima, M.~Larkin, G.~M. Luke, J.~Merrin,
  O.~Tchernysh{\"{o}}v, Y.~J. Uemura, N.~Ichikawa, M.~Goto, and S.~Uchida,
  Phys. Rev. Lett. {\bfseries 77}, 5421 (1996).

\bibitem{Abrikosov:SPJ12:1961:1243}
A.~A. Abrikosov and L.~P. Gor'kov, Sov. Phys. JETP {\bfseries 12}, 1243 (1961).

\bibitem{Adachi:arXiv:1512.08095:2015}
T.~Adachi, A.~Takahashi, K.~M. Suzuki, M.~A. Baqiya, T.~Konno, T.~Takamatsu,
  M.~Kato, I.~Watanabe, A.~Koda, M.~Miyazaki, R.~Kadono, and Y.~Koike, arXiv: .

\bibitem{Takamatsu:PP58:2014:46}
T.~Takamatsu, M.~Kato, T.~Noji, and Y.~Koike, Phys. Procedia {\bfseries 58}, 46
  (2014).

\bibitem{Yamamoto:JPSJ85:2016:024708}
M.~Yamamoto, Y.~Kohori, H.~Fukazawa, A.~Takahashi, T.~Ohgi, T.~Adachi, and
  Y.~Koike, J. Phys. Soc. Jpn. {\bfseries 85}, 024708 (2016).

\end{thebibliography}

\newpage

%%%%%%%%%%%%%%%%%%%%%%%%%%%%%%%%%%%%%%%%%%%%%%%%%%%%%%%%%%%%

%%%%%%%%%%%%%%%%%%%%%%%%%%%%%%%%%%%%%%%%%%%%%%%%%%%%%%%%%%%%
\end{document}